\documentstyle[aps,epsfig]{revtex}

\newcommand{\boldk}{{\bf k}}
\newcommand{\boldr}{{\bf r}}

\begin{document}


\title{Dynamics of ferromagnetic spherical spin models with power law interactions: exact solution}

\author{Sergio A. Cannas$^1$\cite{auth1}, Daniel A.Stariolo$^{2}$
\footnote{Regular Associate of the Abdus Salam International Center for
Theoretical Physics, Strada Costiera 11, 34014 Trieste, Italy} and Francisco A.Tamarit$^1$\cite{auth1}}

\address{$1-$ Facultad de Matem\'atica, Astronom\'\i a y F\'\i sica,
Universidad Nacional de C\'ordoba, Ciudad Universitaria, 5000
C\'ordoba, Argentina.\\
$2-$Instituto de F\'\i sica, Universidade Federal do Rio Grande do Sul,
CP 15051, 91501-970 Porto Alegre RS, Brazil.\\
E-mails: {\tt cannas@famaf.unc.edu.ar,stariolo@if.ufrgs.br,tamarit@famaf.unc.edu.ar}}
\maketitle

\begin{abstract}
We solve the Langevin dynamics of d-dimensional  ferromagnetic spherical models with interactions
that decay with distance as $r^{-(d+\sigma)}$.
The long time dynamics of correlations
and responses are studied in detail in the different dynamical regimes and the
validity of fluctuation-dissipation relations (or its violation) are shown.  In particular, we show
that the fluctuation-dissipation ratio $X(t+t_w,t_w)$ is assymptotically a fucntion only of the waiting
time $t_w$ in the aging regime and that $X\rightarrow 0$ as $t_w\rightarrow\infty$.
The results are valid in any finite dimension $d$ and for
$0 < \sigma < 2$ where short range behavior is recovered.

We also solve the $T=0$ Cahn-Hilliard dynamics of this  model (conserved order parameter).
An analysis of the multiscaling behavior of the autocorrelation function is presented.

\vspace{1cm}

\end{abstract}


\section{Introduction}
Systems with long range interactions are very common in nature. Some important
examples are dipolar systems in which the interaction decays with distance as
$1/r^3$\cite{binder,Toloza,dipolar}; charged systems with Coulomb interactions $\propto 1/r$\cite{chayes}; spin glasses
characterized by RKKY interactions\cite{binder2};
block copolymers\cite{ohta,bahiana} and
models of structural glasses\cite{kivelson}, to name a few. In spite of their
ubiquity these kind of systems receive normally much less attention than short
range ones, perhaps because of its grater analytical complexity. Power law
decaying interactions interpolate between the much studied, although not
realistic, mean field limit and the strictly local nearest neighbour
interactions. Although short range interactions are assumed to be dominant in
a great variety of systems, this may not be so in many others, spin glasses
being a known and controvertial example.

In pure systems, the dynamics after a quench in
temperature from the high temperature phase to the low ordered one proceeds
by a slow
coarsening of domains characteristic of the low energy excitations of the
system. This coarsening process is characterized by growth laws which show
how the typical lenght scales associated with the domains grow with time.
Typical growing laws are $l(t)\propto t^{1/z}$ with $z$ a dynamic exponent
characteristic of the universality class of the system. Order parameters,
correlation and response functions also show typical scaling behaviour.
Exponents and scaling functions also depend on the dynamics being with
conserved or non conserved order parameter and also on the nature of the order
parameter. From here on we will discuss only
results for the vector order parameter case ($n > 1$). In the case of long range forces
decaying as $r^{-(d+\sigma)}$, a calculation of the $n\rightarrow \infty$ limit
of the O(n) model\cite{hayakawa} gives a growth law
$l(t) \propto t^{1/\sigma}$,
for non conserved order parameter. In the conserved case there appears the
phenomenon of ``multiscaling'' with two characteristic lenght scales:
$l_1 \propto t^{1/(2+\sigma)}$ and $l_2 \propto (t/\ln{t})^{1/(2+\sigma)}$.
Here we rederive these results working diretly on the spherical model at
finite temperature and for general spatial dimension $d$. We will also
derive the scaling forms for the structure factor, two times autocorrelations
and associated responses. In recent years much attention have been devoted to
a possible extension of the fluctuation dissipation relations to the case of
non equilibrium dynamics. It turns out that, in many cases, the fluctuation-
dissipation theorem (FDT) may be
generalized by introducing the concept of time dependent ``effective
temperatures''\cite{leticia}. These are difficult to obtain both analytically
and also to be measured in experiments. The simplicity of the spherical model
permits us to analyze the character of the violation of the FDT during the
coarsening process and to obtain explicitly the effective temperature.

It is known that the ferromagnetic spherical model with long range interactions
has a phase transition in all dimensions $d$ (contrary to the short range
model which only has transition for $d>2$)\cite{joyce}.
In $d=1$ a ferromagnetic phase
is present at finite temperature provided that $0<\sigma<1$ and also in $d=2$
provided
that $0<\sigma<2$. Our results concerning the ordering dynamics after a quench
from the high temperature phase to the low temperature one are valid for every $\sigma$
such that $0<\sigma<s$, where $s=d$ for $d \leq 2$ and $s=2$ for $d>2$. The case of $\sigma=2$ will be excluded as the systems
in $d=1$ and $d=2$ have no phase transitions in this case. For $d>2$ and
$\sigma >2$ the system has the same critical properties of the model with short range interactions.

\section{The model}

We consider spherical spin models consisting of
$N$ continuous spins $s_i(t)$ ($i=1,\ldots,N$) which satisfy for all times $t$ the spherical constraint

\begin{equation}
\sum_{i=1}^N s_i^2(t) = N.
\end{equation}

\noindent The Hamiltonian of the system is given by

\begin{equation}
{\cal H} = - \sum_{(i,j)} J(\boldr_{ij}) s_i s_j
\label{H(s)}
\end{equation}

\noindent where the sum runs over all distinct pairs $(i,j)$ of spins of a $d$-dimensional hypercubic lattice  and $\boldr_{ij}$ is the distance vector between sites $i$ and $j$. The interactions $J(\boldr_{ij})$ decay as a power law of the distance between a pair
of spins in the following way:

\begin{equation}
J(\boldr_{ij}) = J_0 \frac{r_{ij}^{-(d+\sigma)}}{\sum_j' r_{ij}^{-(d+\sigma)}} \;\;\;\; for \;\;\;\ i \neq j
\end{equation}

\noindent where $\sum'_j$ runs over all sites $j\neq i$; $d$ is the space dimension, $\sigma > 0$, $J_0>0$ and $J(0)=0$.
Let us now introduce the Fourier transform of the spin variables $s_i(t)$

\begin{equation}
s_{\bf k}(t) = \frac{1}{\sqrt{N}} \sum_j s_j(t) e^{i{\bf k\cdot r}_j}
\end{equation}

\noindent with $s_{\bf -k}(t) = s_{\bf k}^{\ast}(t)$. Considering periodic boundary conditions the wave vector is
${\bf k}=(k_1,\ldots,k_d)$ with $k_i=2\pi n_i/L, n_i=0,\pm1,\pm2,\ldots,\pm(L/2-1),
\pm(L/2)$ ($i=1,\ldots,d$) and $N=L^d$. Now defining

\begin{equation}
J({\bf k}) \equiv \sum_{\bf r} J({\bf r}) e^{i{\bf k\cdot r}}
\end{equation}

\noindent the Hamiltonian (\ref{H(s)})  is diagonalized,

\begin{equation}
{\cal H} = - \sum_{\bf k} J({\bf k}) |s_{\bf k}(t)|^2
\end{equation}

\noindent with \cite{joyce,chen}

\begin{equation}
J({\bf k}) = J_0 \frac{\sum_{\bf l}' |{\bf l}|^{-(d+\sigma)}
\cos{({\bf k\cdot l})}}
{\sum_{\bf l}' |{\bf l}|^{-(d+\sigma)}},
\label{J(k)}
\end{equation}

\noindent where the sum $\sum_{\bf l}'$ is over all lattice vectors ${\bf l}\neq 0$. 
The critical temperature of this model is given by \cite{joyce}

\begin{equation}
\beta_c J_0 = \frac{1}{N} \sum_{\bf k} \frac{1}{1-J({\bf k})/J_0}
\label{betac}
\end{equation}

In the long wavelength scaling limit $k << 1$ Eq.(\ref{J(k)}) behaves as \cite{joyce2},

\begin{equation}
J({\bf k}) \approx J_0 (1-Ck^{\sigma}+O(k^2)) \label{Jk}
\end{equation}

\noindent  where 
\begin{eqnarray}
C & = & \frac{I_{d,\sigma}}{v_a \Omega_{d,\sigma}(\cal{L})} \\
I_{d,\sigma} & = & \frac{2^{1-\sigma} \pi^{d/2} \Gamma(1-\sigma/2)}
                    {\sigma \Gamma(d/2+\sigma/2)}
\end{eqnarray}
$v_a$ is the volume of a unit cell in a d-dimensional Bravais lattice $\cal{L}$ and
$\Omega_{d,\sigma}(\cal{L}) = \sum_{\bf l}' |{\bf l}|^{-(d+\sigma)}$.
This approximation is valid for $0 < \sigma < 2$. At $\sigma=2$ logaritmic
corrections are present in $d=2$. In $d$ dimensions and for $\sigma > 2$ the
leading term in the above expansion is $O(k^2)$ and so the critical properties
of the model are those corresponding to short range interactions. The
dynamical properties in this case are also well known (see e.g.
\cite{newman,horner}). The analysis we present in this work can be extended to the mean filed regime
$-d \leq \sigma < 0$. However,  several works on related models \cite{Cannas,Tamarit,Vollmayr} suggest that  the critical properties (both statical and dynamical) in the whole interval are the same as those for the case $\sigma=-d$ (mean field). The dynamics for this last case can be almost trivially solved, showing no coarsening effects (for instance, the two-times autocorrelation function always decays exponentially with the difference of times). Hence, we will not discuss this further in this paper.

We will concentrate on the analysis of the correlation and response functions. The two-time autocorrelation function is defined as

\begin{eqnarray}
C(t,t') & = & \frac{1}{N}\sum_{\bf r} <s_{\bf r}(t) s_{\bf r}(t')>\\
& = & \frac{1}{N}\sum_{\bf k} C_{\bf k}(t,t')
\end{eqnarray}

\noindent with $C_{\boldk}(t,t')\equiv \langle s_{\boldk}(t) s_{\bf -k}(t')\rangle$.
Note that
$C_{\boldk}(t,t)=\sum_{\boldr} C(\boldr,t) e^{-i\boldk \cdot \boldr}$ is the
dynamic structure factor while $C(\boldr,t) \equiv \langle s_0(t) s_{\boldr}(t)
\rangle$ is the spatial correlation function.  The two-time response function is defined as

\begin{eqnarray}
G(t,t') & \equiv & \frac{1}{N} \sum_{\boldr} \left. {\frac{\delta \langle
s_{\boldr}(t) \rangle}{\delta h_{\boldr}(t')}} \right|_{h=0} \\
& =    & \frac{1}{N} \sum_{\boldk}G_{\boldk}(t,t')
\end{eqnarray}

\noindent with

\begin{equation}
G_{\boldk}(t,t')=\frac{\delta\langle s_{\boldk}(t)\rangle}
{\delta h_{\boldk}(t')}.
\end{equation}

\noindent where $h_{\boldr}(t)$ is an inhomogeneous external magnetic field and

\[ h_{\boldk}(t) = \frac{1}{\sqrt{N}} \sum_{\bf r} h_{\boldr}(t) e^{i{\bf k\cdot r}}. \]

\section{Non conserved order parameter}

In this case we  consider a Langevin dynamics for the spins,
\begin{equation}
\frac{\partial s_i}{\partial t}= -\frac{\delta \cal{H}}{\delta s_i} -
z(t) s_i(t) + \xi_i(t), \label{lang}
\end{equation}
where the Lagrange multiplier $z(t)$ enforces the spherical constraint at
each instant and $\xi$ is a gaussian white noise with moments $<\xi_i(t)>=0$
and $<\xi_i(t)\xi_j(t')>=2T\delta_{ij}\delta(t-t')$ as usual.

The Langevin equation in Fourier space reads,
\begin{equation}
\frac{\partial s_{\bf k}}{\partial t} = (J({\bf k}) - z(t))s_{\bf k}(t) +
\xi_{\bf k}(t) \label{langevin}
\end{equation}

with $<\xi_{\bf k}(t)>=0$ and
$<\xi_{\bf k}(t)\xi_{\bf k'}(t')>=2T\delta_{{\bf k,-k'}}\delta(t-t')$. The
formal solution of (\ref{langevin}) is:

\begin{equation}
s_{\bf k}(t) = s_{\bf k}(0) e^{J({\bf k})t-\int_0^t z(t')dt'} +
\int_0^t e^{J({\bf k})(t-t')-\int_{t'}^t z(t'')dt''}
\xi_{\bf k}(t') dt'  \label{formal}
\end{equation}

From (\ref{formal}) and setting
$t > t'$ we get
\begin{equation}
C_{\boldk}(t,t') = \frac{1}{\sqrt{\Xi(t)\Xi(t')}}
\left[ C_{\boldk}(0,0)e^{J({\bf k})(t+t')} +
2T\int_0^{t'}e^{J({\bf k})(t+t'-2t'')}\Xi(t'')dt''
\right]  \label{Ck}
\end{equation}
where $ C_{\boldk}(0,0)=\sum_{\boldr}C(\boldr,0)e^{-i\boldk \cdot \boldr}$ is
the initial autocorrelation and

\[ \Xi(t) \equiv e^{2\int_0^t z(t') dt'}.     \]

\noindent Again using
(\ref{formal}) we easily obtain

\begin{equation}
G_{\boldk}(t,t')=e^{J({\boldk})(t-t')}\sqrt{\frac{\Xi(t')}{\Xi(t)}}.
\label{resp}
\end{equation}

\noindent  From the spherical constraint $C(t,t)=1$ $ \forall t$ we obtain a selfconsistent Volterra equation for
the function $\Xi(t)$:

\begin{equation}
\Xi(t) = \frac{1}{N} \sum_{\boldk} \left[C_{\boldk}(0,0)e^{2J({\boldk})t} +
2T\int_0^{t}e^{2J({\boldk})(t-t')}\Xi(t')dt' \right] \label{Xi}.
\end{equation}

In order to solve this equation for $\Xi(t)$ we must specify the initial correlation
$C(\boldr,0)$.
For a random initial configuration with magnetization $m_0$ we
can choose  $C(\boldr,0)=\delta_{\boldr,0}+(1-\delta_{\boldr,0})m_0^2$, so
$C_{\boldk}(0,0)=(1-m_0^2)+m_0^2 N\delta_{\boldk,0}$.
We will choose $m_0=0$
which is an interesting case for studying the phase ordering dynamics after
a quench from the disordered to the low temperature ordered phase\cite{bray}.
With this choice $C_{\boldk}(0,0)=1$.

The simplest and most relevant case to
study is a quench to zero temperature. It has been shown that scaling functions
and exponents are the same in the whole low temperature phase and the role of
temperature flutuations only amounts to a renormalization of the
amplitudes\cite{newman}. We will show this results emerging cleanly from the
exact solution of the model at temperatures $T < T_c$.

For $T=0$ we obtain from Eq.(\ref{Xi}) in the thermodynamic limit $N\rightarrow \infty$:

\begin{equation}
\Xi(t) = \frac{1}{N} \sum_{\boldk} e^{2J({\boldk})t}
\rightarrow \frac{1}{(2\pi)^d}\int d\boldk e^{2J({\boldk})t}.
\end{equation}

\noindent Using (\ref{Jk}) and making a change of variables $u=k^{\sigma}t$ this integral
can be explicitely evaluated:

\begin{equation}
\Xi(t) \propto \frac{e^{2J_0t}}{t^{d/\sigma}}\int_0^{2\pi^\sigma t}
u^{\frac{d-\sigma}{\sigma}} e^{-J_0Cu} du
\end{equation}

For $0<\sigma<d$ and $t \gg 1$ the asymptotic behavior is

\begin{equation}
\Xi(t) \sim  A \frac{e^{2J_0t}}{t^{d/\sigma}} \label{xi0}
\end{equation}

\noindent with

\[ A = \frac{\Omega_d}{(2\pi)^d}\left(  \frac{d}{\sigma} \right) 2^{-\frac{d}{\sigma}}
\int_0^\infty u^{\frac{d-\sigma}{\sigma}} e^{-J_0 Cu} du \]

\noindent where $\Omega_d$ is the volume of a d-dimensional hypersphere of unit radius.
Using this result we obtain for the two time autocorrelation:

\begin{eqnarray}
C_{\boldk}(t,t') & = & \frac{e^{J_0(1-Ck^{\sigma})(t+t')}}
{\sqrt{\Xi(t)\Xi(t')}}  \\
& \sim & \frac{1}{A}e^{-J_0Ck^{\sigma}(t+t')}(tt')^{\frac{d}
{2\sigma}}
\end{eqnarray}

In particular we note that the dynamic structure factor

\begin{equation}
C_{\boldk}(t) = \frac{1}{A}e^{J_0Ck^{\sigma}t}t^{\frac{d}{\sigma}}
\label{struct}
\end{equation}

\noindent shows the expected scaling form

\begin{equation}
C_{\boldk}(t) = L^d(t)f(kL(t))
\end{equation}

\noindent with

\begin{equation}
L(t)=t^{1/\sigma},
\end{equation}

\noindent thus recovering the $n\rightarrow \infty$  limit of the O(n) model \cite{hayakawa,bray2}.

It is known that  in the asymptotic coarsening  regime  a ferromagnet  {\it ages}, that is, two time correlation and response functions depend on both times $t$ and $t'$ explicitly through a non trivial scaling form.
In coarsening systems the dependence is of the form $L(t)/L(t')$ ($t,t'\gg 1$), i.e.,
correlations and responses depend on both times through the ratio of the corresponding 
characteristic lenghts. We will analyze the aging dynamics of the present system
with the inclusion of finite temperature fluctuations. In doing so we will see
that two times scaling laws in the aging regime are {\bf exactly} the same,
asymptotically, for every temperature $T < T_c$. At finite temperatures the Volterra equation (\ref{Xi}) can be solved by Laplace transforming $\Xi(t)$:
\begin{equation}
\widetilde{\Xi}(s) = \int_0^{\infty} \Xi(t) e^{-st} \,dt
\end{equation}

Using that $C_{\boldk}(0,0)=1$ the Laplace transform $\widetilde{\Xi}(s)$ of $\Xi(t)$ results:

\begin{equation}
\widetilde{\Xi}(s) = \frac{ \frac{1}{N}\sum_{\boldk}\frac{1}
{s - 2J(\boldk)}}
{ 1 - 2T\frac{1}{N}\sum_{\boldk}\frac{1}
{s - 2J(\boldk)}}   \label{xis}
\end{equation}

In order to obtain the scaling behavior of $\widetilde{\Xi}(s)$ we have to calculate first the function

\begin{equation}
K(s) \equiv \frac{1}{N}\sum_{\boldk}\frac{1}{s - 2J(\boldk)}  \label{Ksum}.
\end{equation}

\noindent In the thermodynamic limit $N\rightarrow\infty$

\begin{equation}
K(s) =\frac{1}{(2\pi)^d}\int \frac{d\boldk}{s - 2J(\boldk)} \label{Kint}
\end{equation}

Introducing (\ref{Jk}) and after some algebra one obtains\cite{joyce,chen} in the scaling regime $k \ll 1$:

\begin{equation}
K_c - K \propto \left\{
\begin{array}{ll}
\epsilon^{\frac{d-\sigma}{\sigma}} & \quad \textrm{if \quad $d/2<\sigma<d$}\\
\epsilon \ln{\epsilon}             & \quad \textrm{if \quad $\sigma = d/2$}\\
\epsilon                           & \quad \textrm{if \quad $0<\sigma<d/2$}
\end{array}
\right.       \label{K-Kc}
\end{equation}

\noindent where $\epsilon = s-2J_0$ and $K_c=K(2J_0)=\beta_c/2$ (see Eq.(\ref{betac})).
From now on
one would proceed by different routes depending on the value of $\sigma$.
Nevertheless it can be shown that the final results for the scaling behaviors of the correlations and responses are the same for all $0 <
\sigma < 2$. Hence, we will only present here the calculations for $d/2<\sigma<d$.

With the help of (\ref{xis}),(\ref{Kint}) and (\ref{K-Kc}) we find

\begin{equation}
\widetilde{\Xi}(s) = \frac{
\frac{\beta_c}{2}-B(s-2J_0)^{\frac{d}{\sigma}-1}
}{
1-\frac{\beta_c}{\beta}+
\frac{2B}{\beta}(s-2J_0)^{\frac{d}{\sigma}-1}
}
\end{equation}

\noindent with $B=A | \Gamma(1-d/\sigma) |$, $\Gamma(x)$ being the gamma function. Working in the low
temperature limit $\beta \gg \beta_c$ and for long times we can antitransform
the above expression obtaining

\begin{equation}
\Xi(t) =  \frac{A}{(1-\beta_c/\beta)^2} \frac{e^{2J_0t}}{t^{d/\sigma}}.
\end{equation}

One can see that the only difference with the $T=0$ expression (\ref{xi0}) is
the factor $(1-\beta_c/\beta)^{-2}$.The time dependence and exponents were not
affected by finite temperature. Now we can go back and calculate the two time
correlation (\ref{Ck}). It is important to note that for the evaluation of
$C_{\boldk}(t,t')$ we need $\Xi(t)$ for all times and not only for long ones.
A detailed analysis of the behavior of $\Xi(t)$ for short as well as long
times shows that its contribution to the long time solution of
$C_{\boldk}(t,t')$ will only be finite for times $t > t_{mic}$, $t_{mic}$
being some microscopic timescale. One then obtains

\begin{equation}
C_{\boldk}(t,t') \sim e^{-J_0Ck^{\sigma}(t+t')}\left[ \frac{(1-\beta/\beta_c)^2}{A}(tt')^{\frac{d}{2\sigma}} +
2Tt'\left(\frac{t}{t'}\right)^{\frac{d}{2\sigma}}
\int_{t_{mic}/t'}^1 e^{2J_0Ck^{\sigma}t'u} u^{-\frac{d}
{\sigma}} du \right] \label{Ck1}
\end{equation}

\noindent  The behaviour of the
autocorrelation depends on the integral which is a function of $k^{\sigma}t'$.

Now we will analyze the two different dynamical regimes $k^{\sigma}t\gg 1$ (fluctuations inside the
domains) and
$k^{\sigma}t\ll 1$ (coarsening regime) and show how this reflects in the
fluctuation-dissipation relations.

\subsection{\bf $k^{\sigma}t' \gg 1$ : bulk fluctuations or quasi-equilibrium behaviour}

In this case the two time autocorrelation (\ref{Ck1}) shows the asymptotic behavior

\begin{equation}
C_{\boldk}(t,t')  \sim  e^{-J_0Ck^{\sigma}(t+t')}\left[
\frac{(1-\beta_c/\beta)^2}{A}
e^{-J_0Ck^{\sigma}t'}(tt')^{\frac{d}{2\sigma}} +
\frac{T}{J_0Ck^{\sigma}}\left(\frac{t}{t'}\right)^{\frac{d}
{2\sigma}}e^{2J_0Ck^{\sigma}t'}\right]
\end{equation}

\noindent and for long times $t \gg t'$ we have that

\begin{equation}
C_{\boldk}(t,t') \sim  \frac{T}{J_0Ck^{\sigma}}\left(\frac{t}{t'}
\right)^{\frac{d}{2\sigma}}e^{-J_0Ck^{\sigma}(t-t')},
\label{Ck2}
\end{equation}

\noindent that is, the dynamics becomes stationary
with exponential decay of correlations. It is simple to obtain the two time
response function (\ref{resp}) which gives

\begin{equation}
G_{\boldk}(t,t') = \left(\frac{t'}{t}\right)^{-\frac{d}{2\sigma}}
e^{-J_0Ck^{\sigma}(t-t')}.
\label{resp2}
\end{equation}

\noindent Comparing Eqs.(\ref{Ck2}) and (\ref{resp2}) we see that the  Fluctuation-Dissipation Theorem:

\begin{equation}
TG_{\boldk}(t,t') = \frac{\partial C_{\boldk}(t,t')}{\partial t'} \label{fdt}
\end{equation}

\noindent is obeyed in this regime. That is, short wavelength fluctuations $k \gg 1/L(t)$ reflect the (local) equilibrium inside the domains.
We now analyze the more interesting coarsening regime.

\subsection{\bf $k^{\sigma}t' \ll 1$ : coarsening, non equilibrium behaviour}

The autocorrelation behaves in this regime as

\begin{equation}
C_{\boldk}(t,t') \sim e^{-J_0Ck^{\sigma}(t+t')} \left[ \frac{(1-\beta_c/\beta)^2}{A}
(tt')^{\frac{d}{2\sigma}}
+\frac{2T}{\left( 1-\frac{d}{\sigma} \right)} t'\left(\frac{t}{t'}\right)^{\frac{d}{2\sigma}} + {\cal O}(k^\sigma t')\right]
\end{equation}

We see that the only effect of temperature on the structure
factor is a renormalization of the amplitude:

\begin{equation}
C_{\boldk}(t) = \frac{(1-\beta_c/\beta)^2}{A} t^{d/\sigma}e^{-2J_0Ck^{\sigma}t}.
\end{equation}

\noindent This has the same scaling as the zero temperature limit (\ref{struct}).

In order to analyze the two time scalings in the aging regime we integrate
out all modes $\boldk$ in the autocorrelation:

\begin{equation}
C(t,t') = \frac{1}{(2\pi)^d}\int d\boldk C_{\boldk}(t,t')
\end{equation}

This gives
\begin{equation}
C(t,t') = \frac{2^{\frac{d}{\sigma}}}{(t+t')^{d/\sigma}}\left[ (1-\beta_c/\beta)^2(tt')^{\frac{d}{2\sigma}}
+\frac{2AT}{\left( 1-\frac{d}{\sigma} \right)} t'\left(\frac{t}{t'}\right)^{\frac{d}{2\sigma}}\right]
\end{equation}

In the aging regime $t \gg t'$ this simplifies to

\begin{eqnarray}
C(t,t') & = & 2^{\frac{d}{\sigma}} (1-\beta_c/\beta)^2 \left(\frac{t'}{t}\right)^{\frac{d}{2\sigma}}\\
& \propto & f\left(\frac{L(t')}{L(t)}\right)
\end{eqnarray}

\noindent which shows the aging scaling typical of coarsening systems with $L(t)=
t^{\frac{1}{\sigma}}$ valid for general vector models. A similar computation for
the response function gives

\begin{equation}
G(t,t') = \frac{2^{\frac{d}{\sigma}} A}{(t-t')^{d/\sigma}} \left(\frac{t}{t'}
\right)^{\frac{d}{2\sigma}}
\end{equation}

It can be readly seen that the Fluctuation-Dissipation Theorem (\ref{fdt}) is
not obeyed in this regime. Nevertheless it can be extended to this non
equilibrium regime defining the so called ``fluctuation-dissipation ratio''
\cite{leticia} as

\begin{equation}
X(t,t') = \frac{TG_{\boldk}(t,t')}{\frac{\partial C_{\boldk}(t,t')}
{\partial t'}}
\end{equation}

This function is particularly interesting in the case of complex systems such
as spin glasses or structural glasses where it can be associated with physical
effective temperatures\cite{leticia}. Its time dependence encodes information
on the time scales structure of the system\cite{Franz}.
In the case of coarsening systems
the response is weak and asymptotically goes to zero, signalling the weakness
of memory effects in these systems\cite{Barrat,dipolar}.
This can be explicitely seen in this model where at long times
\begin{equation}
X(t,t') \propto (t')^{1-\frac{d}{\sigma}} \rightarrow 0  \label{X}
\end{equation}
We see in the above result that $X(t,t')$ is, asymptotically, a function only
of $t'$, as is observed in models of structural glasses. This implies that,
in particular
for fixed $t'$, the fluctuation-dissipation ratio is a constant. The value of
the exponent $1-\frac{d}{\sigma}$ also shows that the dynamics becomes faster,
and memory effects weaker, as the interactions become more long ranged. This
means that ferromagnetic domain walls move faster as long range interactions
become more important.

\section{Conserved order parameter}

We now consider the continuous limit of the spherical model, that is, 
the spin variables $s_i(t)$ are replaced by a field $s({\bf r},t)$, ${\bf r}$ 
now being a d-dimensional continuous position vector, and where the field is  
subject for all times $t$ to the spherical constraint:

\[ \int d{\bf r} \left[ s({\bf r},t) \right]^2 =V,  \]

\noindent where the integral is carried out over an  hypercube of side $L$ with $V=L^d$. We will restric the analysis to zero temperature. In the case
of conserved order parameter the dynamics is governed by the Cahn-Hilliard
equation:

\begin{equation}
\frac{\partial s({\boldr},t)}{\partial t}= \nabla^2 \frac{\delta \cal{H'}}{\delta
s({\boldr},t)}
\end{equation}

\noindent where the Hamiltonian takes now the form

\begin{equation}
{\cal H'} = - \int d{\bf r} \int d{\bf r'} J\left( |{\bf r}-{\bf r'}| \right) s({\bf r},t) s({\bf r'},t) + z(t) \int d{\bf r} \left[ s({\bf r},t) \right]^2.
\end{equation}

\noindent Transforming Fourier we arrive at

\begin{equation}
\frac{\partial s_{\boldk}}{\partial t}=-k^2 \frac{\delta \cal{H}}{\delta
s_{\boldk}}=k^2 \left\{ [J(k)-z(t)] s_{\boldk} \right\}
\label{sk2}
\end{equation}

\noindent where the asymptotic form (\ref{Jk}) is assumed for $J(k)$ .
The formal solution of Eq.(\ref{sk2}) is now:

\begin{equation}
s_{\boldk}(t) = s_{\boldk}(0) e^{k^2 J({\bf k})t-k^2 \int_0^t z(t')dt'},
\label{solk}
\end{equation}

\noindent and the structure factor reads:

\begin{equation}
C_{\boldk}(t) = C_{\boldk}(0) e^{2k^2J(k)t - 2k^2g(t)}
\label{Ck3}
\end{equation}

\noindent with $g(t)\equiv\int_0^t z(t')dt'$ and considering again $C_{\boldk}(0)=1$. Following \cite{hayakawa} we will obtain the
scaling form of the structure factor by making the reasonable assumption that
$C_{\boldk}
(t)$ will present a maximum as a function of $k$ at some $k_m(t)$ and at long
times it will evolve into a Bragg peak. From the spherical constraint $(2\pi)^{-d} \int d{\bf k} C_k(t) =1$ it follows that $C_{k_m}(t)$ scales with $k_m$ as

\begin{equation}
C_{k_m}(t) \propto k_m^{-d} \label{scal}.
\label{Ck4}
\end{equation}

Maximizing $C_k(t)$ we obtain

\begin{equation}
k_m^{\sigma} = \frac{2}{2+\sigma}\frac{\left[J_0t-g(t)\right]}{J_0Ct}
\label{kms}
\end{equation}

\noindent From Eqs.(\ref{Ck3}), (\ref{Ck4}) and (\ref{kms}) we get

\begin{equation}
\frac{k_m^{2+\sigma}}{\ln{k_m}} = -\frac{d}{J_0 C \sigma t}
\end{equation}

\noindent whose asymptotic solution for long times is

\begin{equation}
k_m \approx \left[ \frac{d}{J_0C\sigma(2+\sigma)}\frac{\ln t}{t} \right]^
{\frac{1}{2+\sigma}}
\end{equation}

We can now reconstruct the structure factor obtaining finally:
\begin{equation}
C_k(t) = \left[ l^d(t) \right]^{\phi(k/k_m)}
\end{equation}

\noindent in which

\begin{equation}
l(t) = t^{\frac{1}{2+\sigma}}
\end{equation}

\noindent and the scaling function

\begin{equation}
\phi(x) = \frac{(2+\sigma)}{\sigma} x^2 - \frac{2}{\sigma} x^{2+\sigma}
\end{equation}

As  in the short range case, the phenomenon of ``multiscaling'' is
also present with long range interactions. In our case two characteristic
lenght scales show up:

\begin{equation}
l(t) \propto t^{1/z} \hspace{2cm} with \hspace{0.5cm}z = 2 + \sigma
\end{equation}

\noindent and

\begin{equation}
k_m^{-1}(t) \propto \left( \frac{t}{\ln t} \right)^{1/z}
\end{equation}

Interestingly, this second lenght scale grows more slowly than $t$ and will
produce a particular scaling form of the two time autocorrelations called
``sub-aging''. This behaviour has been studied in detail recently in
\cite{berthier} for the O(n) model with short range interactions in the
$n\rightarrow \infty$ limit. Here we generalize the result to systems with
power law interactions.

Now it is straitforward to write the solution to (\ref{solk}) at long times:

\begin{equation}
s_{\boldk}(t) = s_{\boldk}(0) \exp{\left\{ -k^{2+\sigma}J_0 C t + k^2 \left[
\frac{d}{2\sigma}\left(\frac{(2+\sigma)}{2}J_0 C t \right)^
{2/\sigma} \ln t \right]^{\frac{\sigma}{2+\sigma}} \right\}}
\end{equation}

From this result we obtain the two time autocorrelation following the
steps done in the non conserved case.
The first important point to note is that, similar to what happens
with short range interactions, the autocorrelation $C(t'+\tau,t')$ relaxes
completely in a time $\tau \sim t'$:

\begin{equation}
C(2t',t') \propto \left( t' \right)^{\frac{d}{2+\sigma}(f-1)} \rightarrow 0
\end{equation}
where
\begin{equation}
f = \left(\frac{2^{\frac{2}{2+\sigma}}+1}{3}\right)^{2/\sigma}
\frac{2^{\frac{2}{2+\sigma}}+1}{2} < 1
\end{equation}
and we have set $J_0 C =1$.
So the dynamics is considerably faster than in the non conserved case.
In the regime when $\tau \ll t'$ we obtain:

\begin{equation}
C(t'+\tau,t') \propto (t')^{ \frac{d}{2+\sigma}
\left[ \left( \frac{2}{\sigma} \right)^{2/\sigma}-1 \right] }
\exp{ \left\{ -\left( \frac{2}{\sigma} \right)^{2/\sigma}
\left( \frac{d}{2+\sigma} \right)
\frac{\ln{t'}}{(2+\sigma)\sigma^2}
\left( \frac{\tau}{t'} \right)^2 \right\} }
\end{equation}

It is interesting to compare this result with the corresponding one for short
range interactions \cite{berthier}. First of all, if $\sigma \neq 2$, the
correlation decays algebraically with $t'$, as noted above. For fixed $t'$
the relaxation goes as

\begin{equation}
C(t'+\tau,t') \propto e^{-b\left(\frac{\tau}{t_r}\right)^2}
\end{equation}

\noindent with $b=\left(\frac{2}{\sigma}\right)^{2/\sigma}
\left( \frac{d}{2+\sigma}\right)\frac{1}{(2+\sigma)\sigma^2}$ and

\begin{equation}
t_r = \frac{t'}{\sqrt{\ln t'}}
\end{equation}

This relaxation time is smaller than $t'$ and the particular aging dynamics is
called ``sub-aging''. The same scaling is observed in the short range case
\cite{berthier}.

Finally we want to point out that, although these results are interesting
{\it per se} as show complex behaviour
emerging from a closed analytic solution of the model, it is important to
note that the presence of multiscaling in dynamics with conserved order
parameter is limited to the spherical limit of general $n$-vector models. Simple
scaling is recovered when corrections to order $1/n$ are considered\cite{bray3}.

\section{Conclusions}

Summarizing, we presented the exact solution for the dynamics of the ferromagnetic spherical model with power law decaying interactions in arbitrary dimension, after a quench from infinite temperature into the ordered phase.  In the case of non-conserved order parameter we analyzed the Langevin dynamics of the model, obtaining the exact long time scaling form for correlations and responses at finite temperature. In the case of conserved order parameter we analyzed a microscopic version of the Cahn-Hilliard dynamics of the model at zero temperature, obtaining the long time scaling form of the two time autocorrelation function. These exact results allow us to check at a microscopic level several scaling hypotesis about coarsening dynamics in pure systems, which are currently derived at a coarse grained level through phenomenologycal Landau-Ginzburg and Cahn-Hilliard equations. In particular, we recovered some know results  about the scaling behavior of the $n\rightarrow\infty$ limit of the $O(n)$ model, by working directly on the spherical model. This allowed us to extend those results by obtainig explicit scaling forms for several correlation and response functions.

One new result worth to mention is that of the Fluctuation-Dissipation ratio Eq.(\ref{X}). This behavior has been observed numerically  in several non disordered systems \cite{dipolar,Barrat} and it has been conjectured to be characteristic of systems which do not present replica symmetry breaking \cite{Franz},
i.e. systems with a single pure state. Up to now, the present result is one of the few exact solutions available confirming that conjecture.

A natural and interesting extension of this work is the study of the dynamics
of frustrated systems without disorder, like models with competing short-range
ferromagnetic and long-range antiferromagnetic (e.g., dipolar) interactions,
in which simulations have shown  a  rich dynamical behavior
\cite{Toloza,dipolar}. In fact, there is a whole class of different physical
systems with these characteristic competing interactions, two examples being
charged systems with weak coulomb interactions\cite{chayes}
and a proposed model for the behaviour of structural glasses\cite{kivelson}.

\vspace{2cm}
{\bf Acknowledgments:} D.A.S. wish to thank the Abdus Salam ICTP for warm hospitality,
where part of this work was completed.
This work is partly supported by the following agencies:  CNPq (Brazil), CONICET (Argentina), CONICOR (C\'ordoba, Argentina) and Secretar\'\i a de Ciencia y Tecnolog\'\i a de la Unversidad Nacional de C\'ordoba (C\'ordoba, Argentina).


\begin{thebibliography}{10}
\bibitem[+]{auth1} {Member of the National Research Council, CONICET 
 (Argentina)}
\bibitem{binder}R. Kretschmer and K. Binder, Z. Phys. B {\bf 34} (1979) 375.
\bibitem{Toloza}J. H. Toloza, F. A. Tamarit and S. A. Cannas,
Phys. Rev. B {\bf 58}, R8885 (1998).
\bibitem{dipolar}D.A. Stariolo and S.A.Cannas, Phys. Rev. B  {\bf 60} (1999)
3013.
\bibitem{chayes}L. Chayes, V.J.Emery, S.A.Kivelson, Z.Nussinov and G.Tarjus,
Physica A{\bf 225} (1996) 129.
\bibitem{binder2}K. Binder and A.P. Young, Rev. Mod. Phys. {\bf 58}, (1986)
801.
\bibitem{ohta}T. Ohta and K. Kawasaki, Macromolecules {\bf 19},  (1986) 2621.
\bibitem{bahiana}M. Bahiana and Y. Oono, Phys. Rev. A {\bf 41}, (1990) 6763.
\bibitem{kivelson}D. Kivelson and G. Tarjus, Phil. Mag. B{\bf 77}, (1998) 245.
\bibitem{hayakawa}H. Hayakawa, Z. R\'acz and T. Tsuzuki, Phys. Rev. E {\bf 47}
(1993) 1499.
\bibitem{leticia}L.F. Cugliandolo, J. Kurchan and L. Peliti, Phys. Rev. E
{\bf 55} (1997) 3898.
\bibitem{joyce}G.S. Joyce, Phys. Rev. {\bf 146} (1966) 349.
\bibitem{joyce2}G.S. Joyce, in {\it Phase Transitions and Critical Phenomena}, 
C. Domb and M. S. Green editors, vol 2, (Academic Press, London, 1972) 375.
\bibitem{newman}T.J. Newman and A.J.Bray, J. Phys. A{\bf 23} (1990) 4491.
\bibitem{horner}W. Zippold, R. K\"uhn and H. Horner, preprint cond-mat/9904329.
\bibitem{chen}Y. Chen, S. Guo, Z. Li and A. Ye, preprint cond-mat/9911198.
\bibitem{Cannas} S. A. Cannas and F. A. Tamarit, Phys. Rev. B {\bf 54}, (1996) 
R12661;  S. A. Cannas, A. C. N. de Magalh\~aes and F. A. Tamarit, Phys. Rev. B 
{\bf 61}, (2000) 11521.
\bibitem{Tamarit} F. A. Tamarit and C. Anteneodo, Phys. Rev. Lett. {\bf 84}, 
(2000) 208 .
\bibitem{Vollmayr} B. P. Vollmayr-Lee and E. Luijten,  preprint cond-mat/0009031.
\bibitem{bray}A.J. Bray, Adv. Phys. {\bf 43}, (1994) 357.
\bibitem{bray2}A.J. Bray, Phys. Rev. E {\bf 47}, (1993) 3191.
\bibitem{berthier}L. Berthier, preprint cond-mat/0003122.
\bibitem{bray3}A. Bray and K. Humayun, Phys. Rev. Lett. {\bf 68}, (1992) 1559.
\bibitem{Barrat} A. Barrat, Phys. Rev. E {\bf 57}, (1998) 3629 .
\bibitem{Franz} S. Franz, M. M\'ezard, G. Parisi and L. Peliti, Phys. Rev. 
Lett. {\bf 81}, (1998) 1758.

\end{thebibliography}
\end{document}